\documentclass[twocolumn,showpacs,prl]{revtex4-2}
\usepackage{graphicx,amsmath,mathrsfs}
\usepackage{amssymb}
\usepackage{amsthm,multirow}
\usepackage{bm,bbm}
\usepackage{subfigure,xcolor}   

\usepackage{color}

\usepackage{tikz}
\usetikzlibrary{shapes.geometric}

\def\bc{\begin{center}}
\def\ec{\end{center}}

\newcommand{\bs}[1]{\boldsymbol{#1}}

\newcommand{\pd}{{\phantom{\dagger}}}
\newcommand{\eps}{\varepsilon}

\newcommand\dhat{{\hat{\bs{\delta}}}}

\newcommand{\im}{{\mathrm{i}}}

\graphicspath{{./}{./figs/}}

\begin{document}
\title{
Nodal semimetals in $d\geq3$ to sharp pseudo-Landau levels by dimensional reduction \\
}
\author{Fabian K\"ohler}
\affiliation{Institut f\"ur Theoretische Physik and W\"urzburg-Dresden Cluster of Excellence ct.qmat, Technische Universit\"at Dresden, 01062 Dresden, Germany}
\author{Matthias Vojta}
\affiliation{Institut f\"ur Theoretische Physik and W\"urzburg-Dresden Cluster of Excellence ct.qmat, Technische Universit\"at Dresden, 01062 Dresden, Germany}

\begin{abstract}
Nonuniform strain applied to graphene's honeycomb lattice can induce pseudo-Landau levels in the single-particle spectrum. Various generalizations have been put forward, including a particular family of hopping models in $d$ space dimensions.
Here we show that the key ingredient for sharp pseudo-Landau levels in higher dimensions is dimensional reduction. We consider particles moving on a $d$-dimensional hyperdiamond lattice which displays a semimetallic bandstructure, with a $(d-2)$-dimensional nodal manifold. By applying a suitable strain pattern, the single-particle spectrum evolves into a sequence of relativistic Landau levels.
We develop and solve the corresponding field theory: Each nodal point effectively generates a Landau-level problem which is strictly two-dimensional to leading order in the applied strain. While the effective pseudovector potential varies across the nodal manifold, the Landau-level spacing does not. Our theory paves the way for strain engineering of single-particle states via dimensional reduction and beyond global minimal coupling.
\end{abstract}

\date{\today}

\maketitle


\paragraph{Introduction. --}
Synthetic gauge fields have become an important tool to engineer phases of matter \cite{aidelsburger_rev}: They enable to realize phenomena associated with orbital magnetic fields for charge-neutral particles; they allow one to create non-Abelian gauge fields; and they can therefore be used to create nontrivial, often topological, states of matter. Methods to achieve synthetic gauge fields include the mechanical deformation of solid-state lattices \cite{si_rev,amorim_rev,naumis_rev}, the rotation of atomic gases \cite{fetter09}, laser-induced Raman transitions of cold atoms in optical lattices \cite{gerbier10}, as well as Floquet engineering in photonic crystals \cite{hafezi}.

In solids, strain engineering is a particularly interesting route which has been used both to design novel physical phenomena and to deliver novel functionalities \cite{si_rev,amorim_rev,naumis_rev}, the latter aspect leading to the emerging field of straintronics \cite{miao21}. One particular goal is to quench the electron kinetic energy, by forming Landau levels or flat bands, such that effects of electron--electron interaction get amplified.
Strain-induced pseudomagnetic fields have been first discussed for carbon nanotubes \cite{suzuura02}. Subsequently, triaxial strain applied to graphene has been argued to induce an approximately homogeneous pseudomagnetic field for low-energy electrons, leading to relativistic pseudo-Landau levels (PLLs) \cite{fogler08,pereira09,guinea10,vozmediano10}. Such PLLs have indeed been observed in strained graphene flakes \cite{levy10}, in graphene wafers grown on a structured substrate \cite{nigge19}, and in artificial molecular structures \cite{manoharan12}. Electronic PLLs induced by non-uniform strain have also been discussed for Weyl semimetals \cite{grushin16,pikulin16,weyl2,ilan20} and superconductors \cite{massarelli17}. For insulating magnets, PLLs of of Néel-state magnons \cite{nayga19} and of Majorana-fermion excitations of spin liquids have been proposed \cite{rachel16a}.
In all these cases, the effect of strain can be understood in terms of a pseudovector potential which influences the orbital motion of elementary excitations via minimal coupling, i.e., mimicking the effect of a physical magnetic field.

Recently, a generalization of strain-induced PLLs known for graphene to arbitrary space dimensions $d$ has been proposed \cite{rachel16b}. This work constructed a family of lattice models with spectral degeneracies inductively but did not provide insights into the physical mechanism leading to PLLs in higher dimensions. Other papers \cite{tang14,uchoa19,lau21} proposed strain schemes for approximately flat bands in $d=3$, but more general insights into their construction are lacking.

\begin{figure}[b]
\centering     
\includegraphics[width=\linewidth]{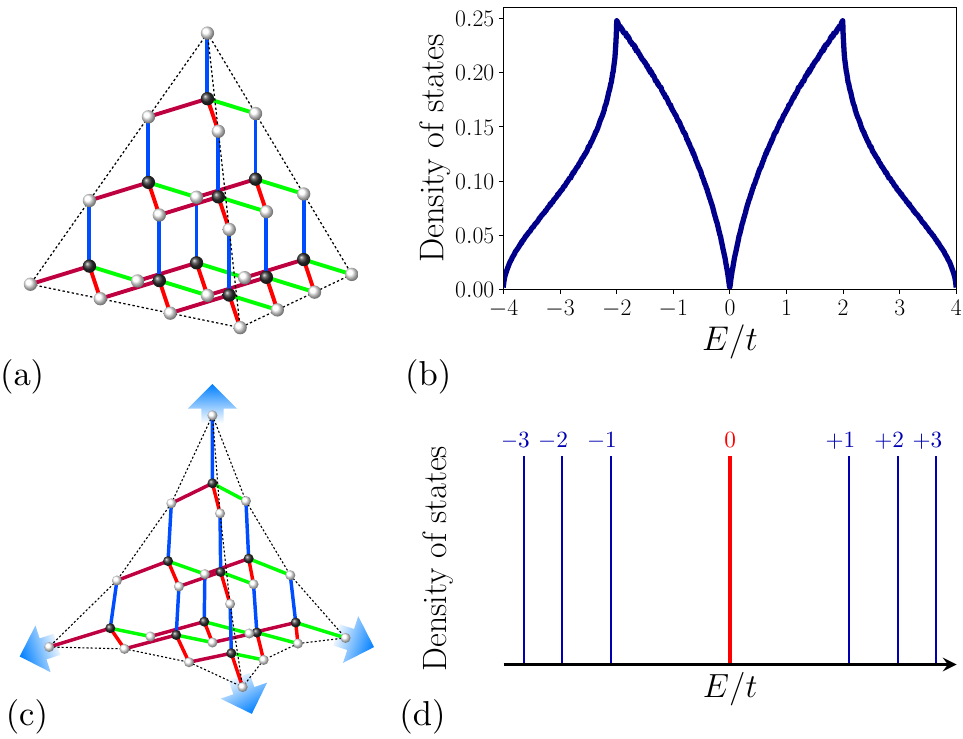}
\caption{
Strain-induced Landau levels in $d=3$:
(a) Diamond lattice with (b) semimetallic density of states.
(c) Diamond lattice under tetraxial strain, resulting in (d) sharp relativistic Landau levels which emerge via dimensional reduction.
}
\label{fig:1}
\end{figure}

It is the purpose of this paper to close this gap. We consider electrons moving on the $d$-dimensional hyperdiamond lattice where the tight-binding dispersion is characterized by a $(d-2)$-dimensional nodal manifold. For $d=3$ we develop and solve the continuum field theory describing the effect of tetraxial strain on the low-energy states (Fig.~\ref{fig:1}), and we generalize this theory to arbitrary $d>3$. Remarkably, the effective theory \textit{cannot} be understood via minimal coupling to a globally defined pseudovector potential. Instead, we show that for $d>2$, dimensional reduction is at play: For each point on the nodal manifold a separate two-dimensional (2D) Dirac theory emerges. All of these Dirac theories generate relativistic PLLs with the \emph{same} Landau-level spacing, and the global excitation spectrum arises from the collection of these two-dimensional Landau-level problems. We also provide a semiclassical picture for $d=3$ which shows how dimensional reduction connects to an ensemble of anisotropic 2D electrons in a \textit{locally} defined pseudomagnetic field.
We discuss experimental realizations and generalizations.

\begin{figure}[!b]
\includegraphics[width=0.9\linewidth]{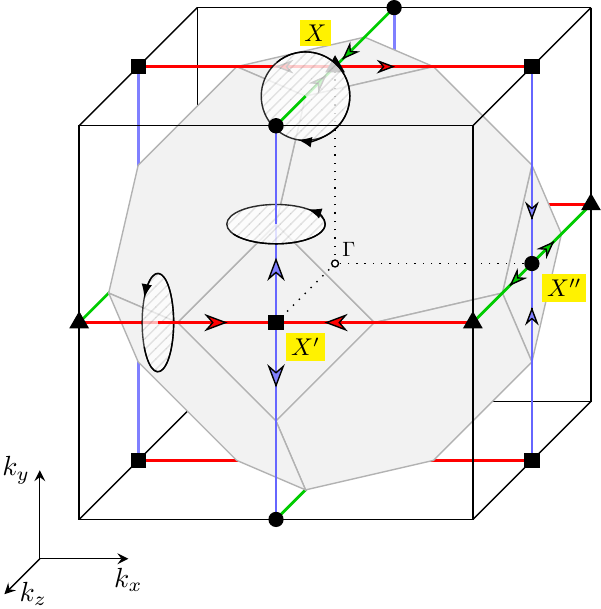}
\caption{
Brillouin zone (shaded) of the diamond lattice, together with the momentum-space network of nodal lines which cross at the high-symmetry $X$ points. Also shown are the directions of the pseudomagnetic field (arrows) emerging under tetraxial strain, together with the corresponding semiclassical trajectories for low-energy electrons (hatched ellipses). The pseudomagnetic field is singular at the $X$ points.
}
\label{fig:FL}
\end{figure}

\paragraph{Hyperdiamond lattice under strain. --}
Motivated by Ref.~\onlinecite{rachel16b}, we consider particles hopping on a hyperdiamond lattice in $d$ space dimensions: This is a bipartite lattice with coordination number $(d+1)$, leading to chain, honeycomb, and diamond lattices in $d=1,2,3$, respectively. The nearest-neighbor vectors $\dhat_j$ connect the center of a $(d+1)$ simplex to each of its vertices, these simplices being line segment, triangle, and tetrahedron in $d=1,2,3$. In the absence of strain we have $\dhat_j^2=a_0^2$, $\dhat_j \cdot \dhat_{j'} = - a_0^2/d$ for $j\not= j'$, and $\sum_{j=1}^{d+1} \dhat_j=\bs 0$, with $a_0$ being the nearest-neighbor distance, which we set to unity in what follows.
We consider a nearest-neighbor hopping Hamiltonian of spinless particles \cite{aniso_note}
\begin{equation}
\label{hh}
\mathcal{H}= - \sum_{\langle ii'\rangle}  t_{ii'} (c_i^\dag\,c_{i'} + {\rm H.c.}) \,.
\end{equation}
In the absence of strain, $t_{ii'}\equiv t$, the momentum-space Bloch Hamiltonian takes the form
$h(\bs k)= t\left( \begin{smallmatrix} 0&f(\bs k)\\ f^\ast(\bs k)&0 \end{smallmatrix} \right)$ in the sublattice basis, resulting in two bands with dispersion
\begin{equation}
\eps(\bs k) = \pm t |f(\bs k)|,~\text{where}~
f(\bs k)=\sum_j \exp[-\im \bs k \cdot \dhat_j].
\end{equation}
As shown in the Supplemental Meterial \cite{suppl}, the function $f(\bs k)$ vanishes on a $(d-2)$-dimensional manifold in momentum space, leading to band-touching Dirac points in $d=2$, nodal lines in $d=3$ (Fig.~\ref{fig:FL}), nodal surfaces in $d=4$ etc. This nodal manifold is characterized by a Berry phase of $\pi$ in any $d\geq 2$; that is, electrons encircling a nodal point along a finite-energy trajectory acquire a $\pi$ phase shift.

Upon distorting the lattice, the electronic hopping amplitudes $t$ get modified because wavefunction overlaps change. Empirically, the hopping amplitudes follow \cite{aniso_note}
\begin{equation}\label{nonlin-hoppings}
t_{ii'} = t_0 \exp\left[ - \beta(|\bs{R}_{ii'}|/a_0 -1)\right],
\end{equation}
where $t_0\equiv t$ is the hopping in the absence of strain and $\bs{R}_{ii'}=\bs{r}_i+\bs{u}_i-\bs{r}_{i'}-\bs{u}_{i'}$ is the distance between sites $i$ and $i'$, where $\bs{u}$ is the displacement field evaluated at the lattice positions $\bs{r}_i$. The factor $\beta$ encodes the strength of electron-lattice coupling: Typical values are of order unity; for graphene, $\beta=3.37$ \cite{peeters13}.

We subject the system to a $d$-dimensional generalization of the triaxial strain introduced for graphene \cite{guinea10}. As will become clear below, this $(d+1)$-axial strain has the property that it generates only a pseudomagnetic field, but no pseudoelectric field. The strain is defined by the displacement field
$\bs u(\bs r)=\frac{C}{2} \sum_j (\bs \dhat_j \cdot \bs r)^2 \bs \dhat_j $, leading to the strain tensor
\begin{equation}\label{d-axial strain}
\underline{\underline{u}} = C \sum_j (\bs \dhat_j \cdot \bs r)(\bs \dhat_j \circ \bs \dhat_j)\,.
\end{equation}
The effect of the strain on the hopping matrix elements is now parametrized by the product $\gamma=\beta C (1-1/d^2)$, with the last factor included for later convenience.
We note that mechanical stability limits the maximum displacement and hence the value of $C$, with the maximum allowed $C$ scaling with inverse linear system size \cite{suppl}. The continuum theory below relies on a Taylor expansion in $C$ and is therefore unaware of this restriction.

\begin{figure}[!bt]
\includegraphics[width=\linewidth]{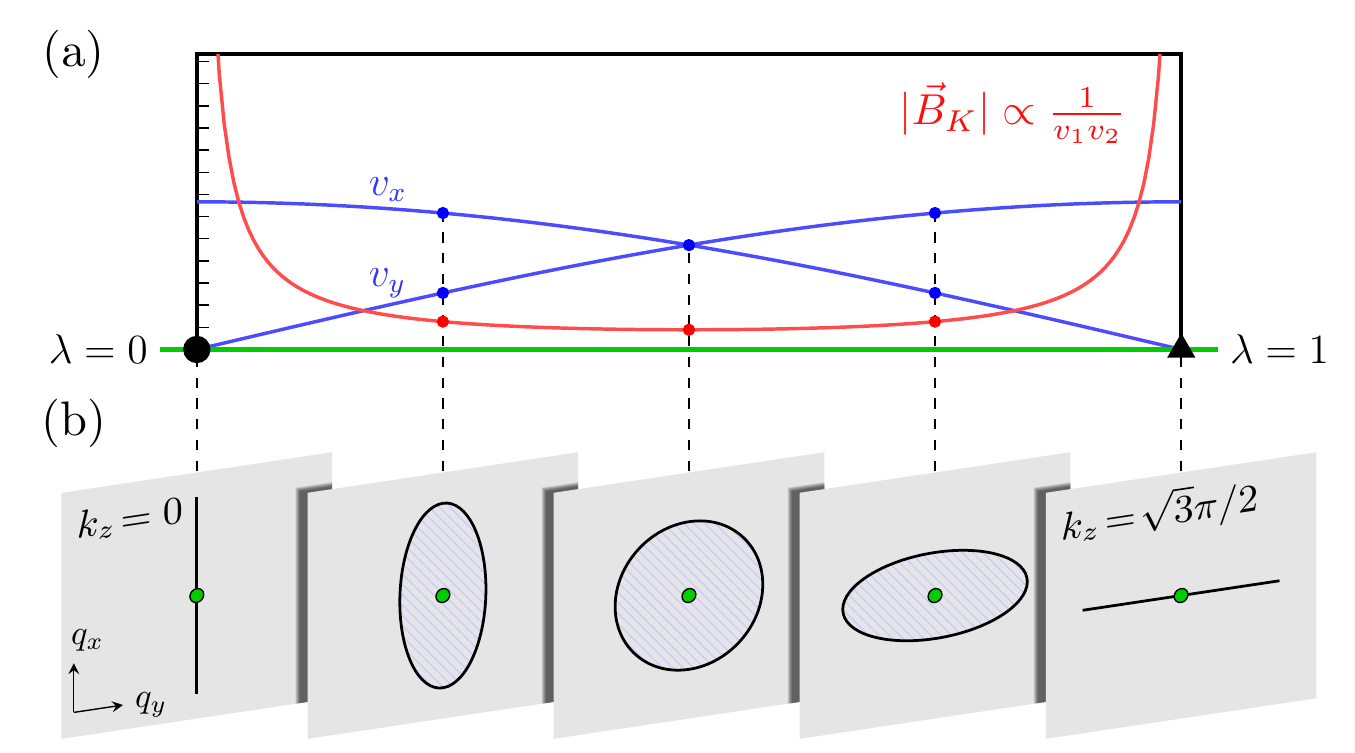}
\caption{
Illustration of dimensional reduction in $d=3$.
(a) Dirac velocities $v_x$ and $v_y$, together with the amplitude of the pseudomagnetic field $|{\bs B}_{\bs K}|$, plotted along a nodal line connecting two $X$ points.
(b) Semiclassical trajectories at selected points along the nodal line in the momentum-space plane perpendicular to the nodal line.
}
\label{fig:slices}
\end{figure}

\paragraph{Continuum theory in $d=3$. --}
To be specific, we demonstrate the continuum treatment for the most relevant case of $d\!=\!3$. The tight-binding model \eqref{hh} on the diamond lattice yields a dispersion featuring three distinct straight nodal lines which connect and cross at different $X$ points \cite{chadi75,takahashi13}; see Fig.~\ref{fig:FL}.
We choose the nodal line along $z$ and expand the Hamiltonian about an arbitrary momentum-space point on the green line segment connectuin a solid cirlce and a solid triangle in Fig.~\ref{fig:FL}, $\bs K = \sqrt{3}\pi/2(1,0,\lambda)$ parametrized by $\lambda \in (0,1)$, to obtain
\begin{equation}
\label{hdirac}
h_{\bs K} (\bs q) = v_x(\lambda)  q_x \sigma_x + v_y(\lambda) q_y \sigma_y + O(q^2),
\end{equation}
where the full momentum is $\bs k=(K_x-q_x,K_y-q_y,K_z-q_z)$ and $\sigma_{x,y}$ are Pauli matrices in the sublattice space. 
Equation~\eqref{hdirac} is a 2D Dirac theory in the plane perpendicular to the nodal line, with velocities $v_x(\lambda) = (4t/\sqrt{3}) \cos(\frac{\pi}{2}\lambda)$, $v_y(\lambda) = (4t/\sqrt{3}) \sin(\frac{\pi}{2} \lambda)$; see Fig.~\ref{fig:slices}(a). One of the velocities vanishes for $\lambda=0,1$, i.e., the $X$ points where two nodal lines cross.
A small-momentum expansion directly at one of the crossing points $X$ yields
\begin{equation}
h_{X}(\bs q) = \frac{4t}{\sqrt{3}} q_x \sigma_x  - \frac{4 t}{3} q_y q_z \sigma_y + O(q^3);
\end{equation}
the full momentum $\bs k$ here is $(\sqrt{3}\pi/2-q_x,-q_y,-q_z)$, and we have included the quadratic piece.

We now proceed to include the effect of strain. We focus on the nodal line described by Eq.~\eqref{hdirac} and incorporate strain-induced changes of the hopping according to Eq.~\eqref{nonlin-hoppings} up to first order in $\gamma$, explicitly using the $d=3$ version of Eq.~\eqref{d-axial strain}, i.e., tetraxial strain with $\bs{u}^\pd_{\rm 3D} =  4C/\sqrt{27} ( yz, zx, xy)$; see Fig.~\ref{fig:1}. Technically, we expand about a fixed momentum-space point $\bs K$ on the nodal line and switch to continuum-limit real space in the plane perpendicular to this line. To leading order, this yields
\begin{align}
\label{hstrain}
h_{\bs K}
&= \left(v_x(\lambda) q_x + \frac{4\gamma}{\sqrt{3}} t \sin(\frac{\pi}{2}\lambda ) y \right)\sigma_x \notag\\
&+ \left(v_y(\lambda) q_y - \frac{4\gamma}{\sqrt{3}} t \cos(\frac{\pi}{2}\lambda ) x \right)\sigma_y
\end{align}
where $q_{x,y} = i\partial_{x,y}$.
For fixed $\lambda$ this is a Dirac theory, now minimally coupled to a pseudovector potential $\bs A_{\bs K} = (A_x,A_y) = \frac{\gamma}{v_x v_y} (v_y^2 y,-v_x^2 x)$; that is, it is obtained from \eqref{hdirac} by the replacement $\bs q \to \bs q+\bs A_{\bs K}$.
The pseudomagnetic field ${\bs B}_{\bs K}$ corresponding to $\bs A_{\bs K}$ is given by $\frac{16 \gamma t^2}{3 v_x v_y}(0,0,1)$, i.e., it is homogeneous in the $x$-$y$ plane, but depends on the position $\lambda$ along the nodal line.
These findings equivalently apply to the other nodal lines where $h_{\bs K}$ \eqref{hstrain} then depends on the coordinates ($\vec r_\perp$) and their conjugate momenta ($\vec q_\perp$) perpendicular to the nodal lines, but not on the momentum $q_\parallel$ along the line. The pseudomagnetic field ${\bs B}_{\bs K}$ is always directed parallel to the nodal line, Fig.~\ref{fig:FL}.


\paragraph{Dimensional reduction. --}
The form of the leading-order Hamiltonian \eqref{hstrain} now implies that each point on a nodal line corresponds to a two-dimensional system of Dirac electrons subjected to a perpendicular homogeneous pseudomagnetic field. Since there is no mixing between these 2D systems to this order, the total low-energy Hamiltonian can be written as
\begin{equation}
\label{hdimred}
\mathcal{H} = \int d{\bs K} h_{\bs K}\,,
\end{equation}
representing a sum of two-dimensional subtheories, with the integral running over the nodal manifold. While the Dirac velocities as well as the strength of the magnetic field vary along each nodal line, and moreover the field direction changes from line to line, the spectrum of pseudo-Landau levels turns out to be \textit{independent} of $\lambda$ and hence constant on the entire nodal manifold \cite{suppl},
\begin{equation} \label{PLL}
E_n^\pm = \pm t \sqrt{n} \sqrt{\frac{32\gamma}{3}},
\end{equation}
with $n$ being the Landau-level index. Hence all nodal points generate the \textit{same} 2D Landau-level spectrum, together resulting in sharp (i.e., nondispersive) pseudo-Landau levels with degeneracy of $L^3$, where $L$ is the linear system size. Equations.~\eqref{hdimred} and \eqref{PLL} represent the central result of this paper.
Parenthetically, we note that the crossing points of two nodal lines require a separate treatment, as the pseudomagnetic field diverges, but the PLL spectrum \eqref{PLL} can be recovered as we will show further down.

As announced above, the spectrum \eqref{PLL} does not (and, in fact, cannot) emerge from the coupling of the particle motion to a globally defined pseudogauge field. Instead, electrons in each two-dimensional submanifold are subject to a different pseudogauge field $\bs A_{\bs K}$; hence minimal coupling applies only \textit{after} dimensional reduction.

We also note that the zeroth PLL resides on one sublattice only \cite{rachel16b}. Similar to the case of graphene, this can be understood as a consequence of the parity anomaly of two-dimensional Dirac electrons \cite{schomerus13}.


\paragraph{Semiclassical motion. --}
One can obtain a semiclassical picture of dimensional reduction by constructing a low-energy wave packet from Bloch states near a nodal momentum $\bs K$. Approximating the motion of the wave packet's center using Hamilton equations yields 2D orbits, parametrized by $s$, in the plane perpendicular to the nodal line, elliptic in both momentum space and real space \cite{suppl},
\begin{equation}
\bar{\bs r}_\lambda = r_0 \Big(v_x(\lambda) \cos s,-v_y(\lambda) \sin s, 0\Big) + {\rm const,}
\end{equation}
as one would expect from a semiclassical 2D Landau problem. The orbits can be centered at an arbitrary real-space point of the 2D slice perpendicular to the nodal line, which recovers a degeneracy of $L^2$ per slice. Using Sommerfeld's phase-space quantization rule, together with the $\pi$ Berry phase of the nodal line, yields exactly the spectrum in Eq.~\eqref{PLL}. Interestingly, this construction works for all points on the nodal lines, including their crossing points $X$ \cite{suppl}.


\paragraph{Beyond leading order. --}
To access higher-order terms in both momentum and strain in the expansion of $h_\lambda$ \eqref{hstrain}, it is advantageous to define the operator $\bs \Gamma = -\im \bs q - \beta C_d \bs r$, where $C_d = C (1-1/d^2)$. This then enables one to expand $f$ in $h_{\bs K} = t\left( \begin{smallmatrix} 0&f_{\bs K}\\ f^\ast_{\bs K} &0 \end{smallmatrix} \right)$ in powers of $\bs \Gamma$; this is facilitated by the fact that hopping and $(d+1)$-axial strain involve terms of the form $\exp[-\im\bs q\cdot\dhat_j]$ and $\exp[-\beta C_d \bs r\cdot\dhat_j]$, respectively, which can be combined using $\bs\Gamma$.
An explicit calculation in $d=3$ \cite{suppl} yields a result which can be cast into the form
\begin{equation}
\label{hsecond}
f_\lambda = \frac{4}{\sqrt{3}} \sqrt{2 \gamma} a_1 + \frac{8 \gamma}{3} a_\lambda a_2 + O(\Gamma^3),
\end{equation}
where $a_1,a_2,a_\lambda$ form a complete set of commuting ladder operators, $[a_i,a_i^\dagger]=1$, with $a_{1,2}$ ($a_\lambda$) encoding momentum and position perpendicular to (along) the nodal line, respectively.

Keeping the first-order term only restores the result above, namely a theory which is local with respect to the nodal line (in the sense that it does not involve the momentum along the nodal line) and displays the spectrum given in Eq.~\eqref{PLL}, with a degeneracy given through $a_2^\dagger a_2$.
The strict dimensional reduction, however, does not hold beyond leading order due to the presence of $a_\lambda$: The physics near a chosen point $\bs K$ is then no longer constrained to the momentum plane perpendicular to the nodal line. Then, a rewriting of the form \eqref{hdimred} is no longer possible, since the degrees of freedom along the nodal line would be overcounted. One may instead resort to finite patches along the nodal line and then use Eq.~\eqref{hsecond} to estimate the broadening of the PLLs; see Supplemental Material for details \cite{suppl}. As the expansion \eqref{hsecond} is controlled by powers of $\sqrt{\gamma}$, the broadening is parametrically small for small $\gamma$, i.e., small strain. Notably, the zeroth PLL remains perfectly degenerate at second order.

Taken together, beyond leading order the individual PLLs acquire a finite width, consistent with lattice calculations \cite{rachel16b} and similar to the case of strained graphene \cite{vozmediano10,peeters13}. The zeroth Landau level is the sharpest and hence best suited for observations and applications.


\paragraph{Generalization to $d>3$. --}
Remarkably, the above leading-order calculation can be generalized to arbitrary $d$, even without precise knowledge of the shape of the nodal manifold. To this end, the calculation is now performed in a coordinate-free fashion. We again expand about a nodal point $\bs K$, now to first order in momentum and strain:
\begin{equation}
f_{\bs K} = \sum_j e^{-\im \bs K \cdot \dhat_j} \dhat_j \cdot (\partial_{\bs r} - \beta C_d \bs r) \equiv b_{\bs K},
\end{equation}
where the operators $\partial_{\bs r}$ and $\bs r$ are defined in the plane perpendicular to the nodal manifold. As already noted above, momentum and strain terms are related by exchanging $\bs q$ and $\beta C_d \bs r$ in $f_{\bs K}$, which allows us to introduce a bosonic operator $b_{\bs K}$.
Noting that $\omega_{\bs K} = [b_{\bs K},b^\dag_{\bs K}]$ corresponds to a harmonic-oscillator energy and accounting for the $2\times 2$ matrix structure of the Bloch Hamiltonian $h_{\bs K}$, we find its spectrum to be $E = \pm t \sqrt{n} \sqrt{\omega_{\bs K}}$. An explicit computation of $\omega_{\bs K}$ shows that it is independent of $\bs K$, $\omega=2\gamma (d+1)^2/d$ \cite{suppl}, such that we can define $a_1 = b_{\bs K}/\sqrt{\omega}$ with $[a_1,a_1^\dagger]=1$. Hence \textit{all} points on the nodal manifold yield the same spectrum of pseudo-Landau levels
\begin{equation}
E_n^\pm = \pm t \sqrt{n} \sqrt{2\gamma (d+1)^2/d},
\end{equation}
which is consistent with Eq.~\eqref{PLL} for $d=3$.

We thus conclude that $(d+1)$-axial strain applied to the hyperdiamond lattice yields sharp PLLs with level scaling as $\pm\sqrt{n}$ for arbitrary dimensions: The specific strain pattern guarantees that the semiclassical motion near any nodal point $\bs K$ is constrained to the normal space of the $(d-2)$-dimensional Fermi manifold. Therefore the problem reduces to a family of two-dimensional Landau-level problems regardless of the dimensionality of the original model, explaining the previous lattice-model findings in Ref.~[\onlinecite{rachel16b}].


\paragraph{Conclusion. --}
We have developed and solved a continuum theory for particles hopping on a $d$-dimensional hyperdiamond lattice under the influence of $(d+1)$-axial strain. This single-particle problem results in sharp PLLs in arbitrary $d$, which emerge for $d>2$ via dimensional reduction: Each momentum-space point along the $(d-2)$-dimensional nodal manifold generates an effectively two-dimensional Landau-level problem, with velocities and pseudo-magnetic field varying across the manifold, but having the same (hence global) Landau-level spectrum.

Our theory opens up a non-trivial way of strain-engineering
single-particle spectra: The PLLs in $d>2$ are not generated via {\em global} minimal coupling, but instead arise in two-dimensional submanifolds of the system. Research on generalizations to other lattice structures is underway; a detailed investigation of the higher-dimensional valley quantum Hall effect is left for future work.
Our work also raises interesting questions about interaction effects in strained hyperdiamond lattices: Will interactions also follow the principle of dimensional reduction? If yes, can one find higher-dimensional cousins of fractional quantum Hall states? If not, do novel forms of fractionalization occur for partially filled PLLs?

Our predictions can be tested in cold-atom experiments. For optical lattices, either spatial variations of beam intensities \cite{pekker15} or density-assisted tunneling \cite{jamotte22} have been proposed to simulate strain effects, with the concrete goal of emulating the physics of strained graphene. The resulting PLLs can be probed using Bragg spectroscopy. We believe these settings can be generalized to $d=3$. Alternatively, a three-dimensional generalization of the inhomogeneous photonic lattices demonstrated in Ref.~[\onlinecite{szameit13}] could be considered.


\acknowledgments

We thank  D. Arovas, D. Ben-Zion, I. Goethel, and S. Rachel for collaborations at an early stage of this work, as well as L. Fritz and A. Lau for discussions.
Financial support from the DFG through SFB 1143 (Project No. 247310070) and the W\"urzburg-Dresden Cluster of Excellence on Complexity and Topology in Quantum Matter -- \textit{ct.qmat} (EXC 2147, Project
No. 390858490)  is gratefully acknowledged.



\begin{thebibliography}{}
\bibitem{aniso_note}
Both the tight-binding model \eqref{hh} and strain dependence of its hopping matrix elements \eqref{nonlin-hoppings} assume spatially isotropic orbitals, i.e., strictly apply to $s$ electrons.
\end{thebibliography}
\end{document}


\title{Supplemental material for:\\
Nodal semimetals in $d\geq3$ to sharp pseudo-Landau levels by dimensional reduction \\
}

\author{Fabian K\"ohler}
\affiliation{Institut f\"ur Theoretische Physik and W\"urzburg-Dresden Cluster of Excellence ct.qmat, Technische Universit\"at Dresden, 01062 Dresden, Germany}
\author{Matthias Vojta}
\affiliation{Institut f\"ur Theoretische Physik and W\"urzburg-Dresden Cluster of Excellence ct.qmat, Technische Universit\"at Dresden, 01062 Dresden, Germany}

\date{\today}

\maketitle


\section{Tight-binding model on a hyper-diamond lattice}

\subsection{Lattice}

\begin{figure}[!b]
\includegraphics[scale=0.35]{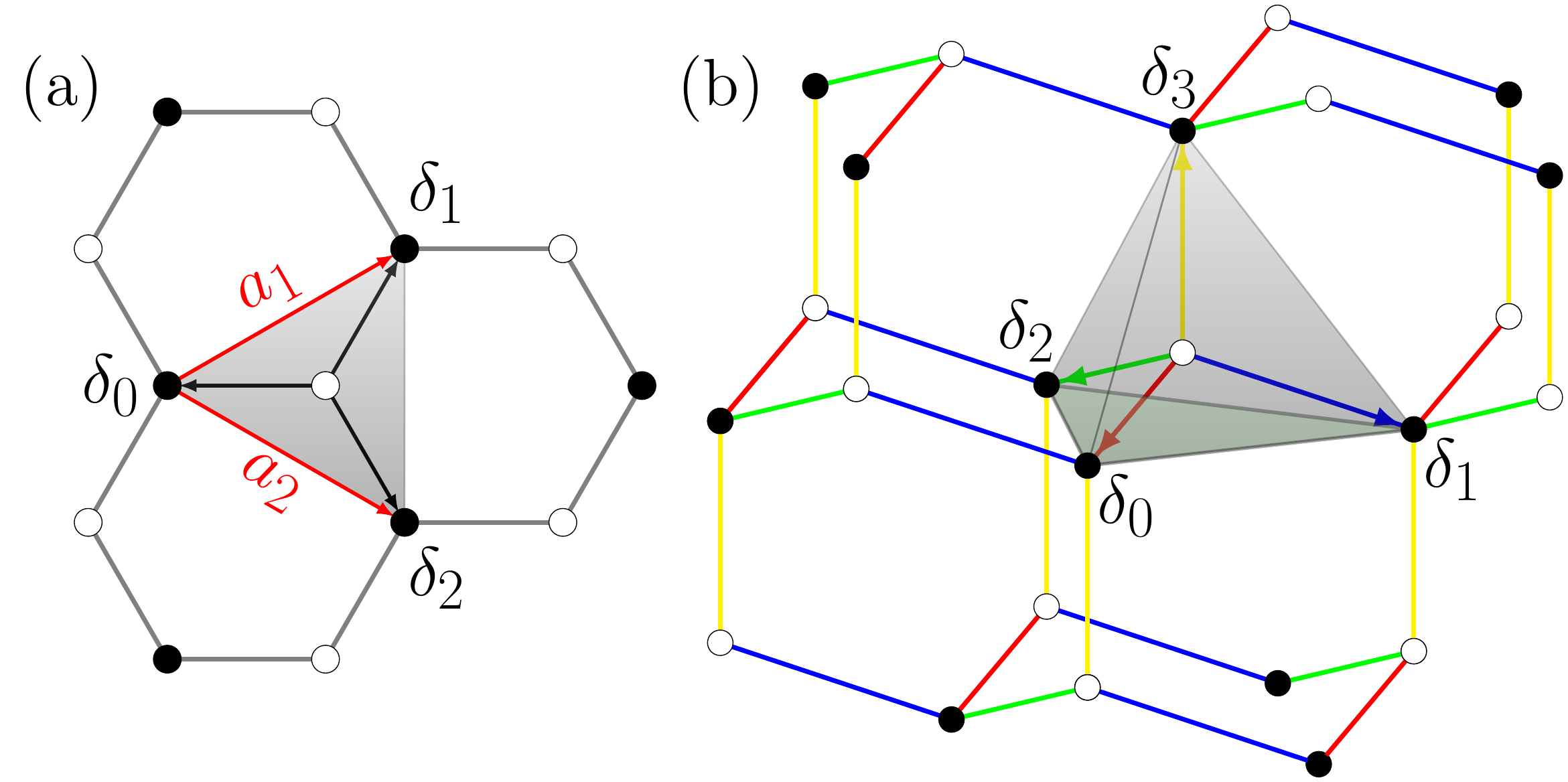}
\caption{
Representatives of the hyper-diamond lattices: (a) honeycomb lattice and (b) diamond lattice. The Bravais-lattice basis vectors are given by $\bs a_j = \dhat_{j+1} - \dhat_1$. The nearest-neighbor vectors fulfill the condition $\dhat_j \cdot \bs a_i = 0$ for $ i \neq j$ so that they can be used for a basis in reciprocal space.
}
\label{fig:honeycomb_lattice}
\end{figure}

The $d$-dimensional hyper-diamond lattices are a family of bipartite lattices in $d$ spatial dimensions, Fig.~\ref{fig:honeycomb_lattice}. Each lattice vertex has a coordination number of $(d+1)$ and the nearest-neighbor vectors $\dhat_j$ from a $d$-simplex centered at the lattice vertex \cite{rachel16b, Kimura10}:
\begin{flalign}
\sum_{j=1}^{d+1} \dhat_j = 0,\qquad
\dhat_j \cdot\dhat_{j'} = -\frac{a_0^2}{d}, j \neq j';
\end{flalign}
the nearest-neighbor distance $a_0$ will be set to unity in the remainder.
The Bravais vectors of the hyper-diamond lattices are given by pairwise differences of the $\dhat_j$. We will define one commonly used set as $\bs a_j = \bs \delta_{j+1} - \bs \delta_1$. Doing so sets the reciprocal lattice vectors to $\bs G_j = 2\pi d/(d+1) \dhat_{j+1}$.

In $d=1,2,3$ the corresponding lattices are (bipartite) chain, honeycomb lattice, and diamond lattice, respectively \cite{Kimura10}.
For the expressions containing explicit coordinates we choose the nearest-neighbor vectors as
\begin{equation}
\dhat_1 = \frac{1}{2}(1,\sqrt{3}),~
\dhat_2 =\frac{1}{2}(1,-\sqrt{3}),~
\dhat_3 = (-1,0)
\end{equation}
in $d=2$, while for $d=3$ we use
\begin{align}
\dhat_1 &= \frac{1}{\sqrt{3}}(1,1,1),~~~~~\, \dhat_2 = \frac{1}{\sqrt{3}}(-1,-1,1), \notag \\
\dhat_3 &= \frac{1}{\sqrt{3}}(-1,1,-1),~\dhat_4 = \frac{1}{\sqrt{3}}(1,-1,-1).
\end{align}
For $d>3$ the nearest-neighbor vectors can be constructed from their lower-dimensional counterparts \cite{rachel16b}.


\begin{figure}[!tb]
\includegraphics[scale=1]{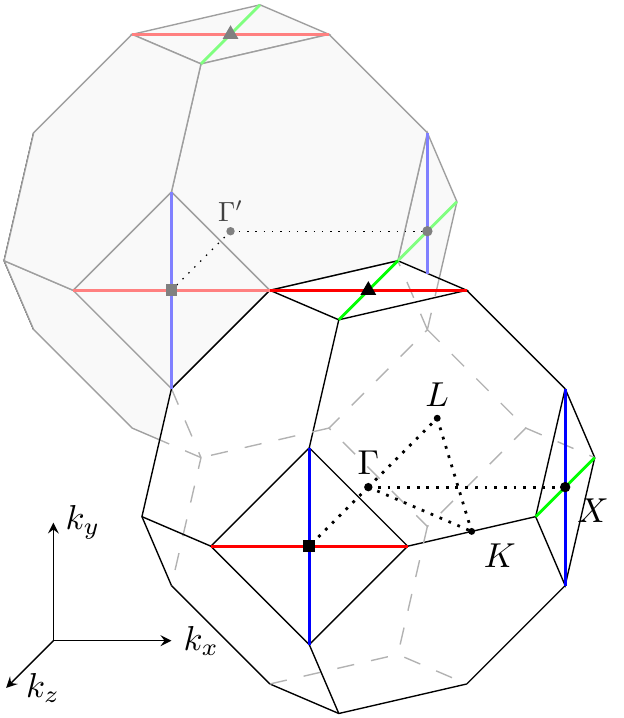}
\caption{
Extended Brillouin zone scheme of the diamond lattice with high-symmetry points, showing the nodal lines (colored) which cross at the $X$ points.
}
\label{fig:BZ}
\end{figure}

\subsection{Nodal manifold}

The nearest-neighbor tight-binding model, Eq.~(1) of the main text, yields a dispersion
\begin{equation} \label{soap}
\eps(\bs k) = \pm t |f(\bs k)|,~
f(\bs k)=\sum_j \exp[-\im \bs k \cdot \dhat_j]
\end{equation}
as usual for bipartite hopping problems. At half-filling, the Fermi level is located at $\eps=0$ due to particle-hole symmetry.
The condition $f=0$ yields the well-known Dirac points at the corners of the Brillouin zone in $d=2$ \cite{Neto09,Masir13} and straight nodal lines in $d=3$ \cite{takahashi13}; the latter are located at the Brillouin zone boundary and cross at the high-symmetry points $X=\sqrt{3}\pi/2(1,0,0)$ (and equivalent), see Fig.~\ref{fig:BZ}. We shall employ a parametrization of the form $\bs K_\lambda = \sqrt{3}\pi/2(1,0,\lambda)$ which we will use for explicit calculations so that one segment between two $X$ points corresponds to a parameter range of $\lambda \in (0,1)$.

\begin{figure}[!tb]
\includegraphics[scale=1]{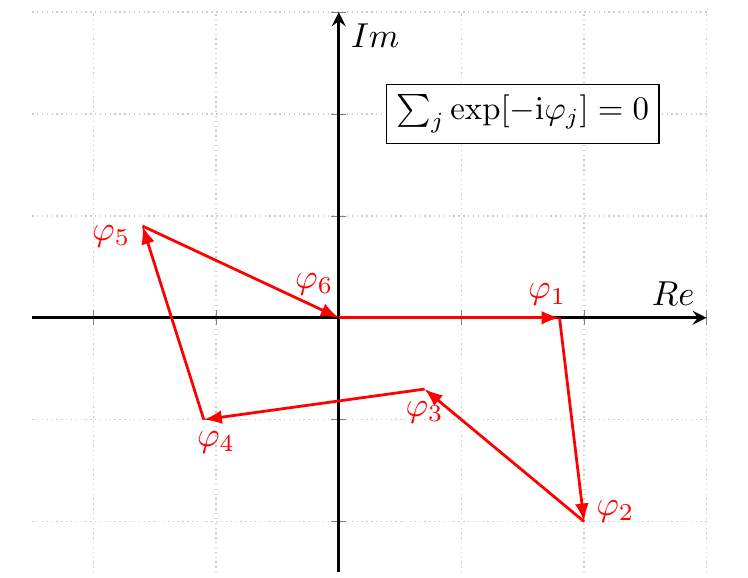}
\caption{
Illustration of a nodal point in $d=5$. The exponentials of projections of $\bs K$ on the nearest-neighbor vectors, $\exp(-\im\varphi_j) = \exp[-\im\bs K\cdot \dhat_j]$ have to sum to zero, therefore the complex vectors $\exp[-\im\varphi_j]$ trace out a polygon in the complex plane.
}
\label{fig:config_space}
\end{figure}

In higher dimensions, $d>3$, we have not succeeded in finding an explicit parametrization for the nodal manifold. We will therefore perform calculations using local coordinate-free operator projections.
However, we can prove that the nodal manifold is $(d-2)$ dimensional as follows: $f(\bs K)$ depends on $d$ independent real parameters, which can be taken either as the components of $\bs K$ or the $\varphi_j = \bs K \cdot \dhat_j$, recalling that $\sum_j \dhat_j = 0$. Requiring $\mathrm{Re}f(\bs K)=\Im f(\bs K)=0$ leaves $(d-2)$ free parameters.
This argument can be made more precise: On the nodal manifold, the $(d+1)$ complex unit vectors $\exp[-\im\varphi_j]$ must sum to zero, i.e., must form a closed polygon in the complex plane, see Fig.~\ref{fig:config_space}. As the global phase of $f(\bs K)$ does not enter, we can perform a ``gauge'' choice by setting $\exp[-\im\varphi_1]=1$. Of the remaining $d$ vectors we can freely choose the first $(d-2)$ as long as the end of their path is inside a disc of radius $2$ around the origin. Inside the disc the $(d-2)$ vectors can be changed arbitrarily, and the remaining two $\exp[-\im\varphi_j]$ are then fixed to close the path. Therefore the nodal manifold has a dimensionality of $(d-2)$.
We conjecture that the nodal manifold is restrained to the surface of the (first) Brillouin zone due to lattice symmetry and particle-hole symmetry for any $d$; we have checked this for selected cuts through the Brillouin zone in $d=4$.

Given the linear band touching, the single-particle density of states therefore scales as $\rho(\omega)\propto|\omega|^1$ in any dimension. We have checked this numerically in $d=3$ (Fig.~1 of the main text) and $d=4$ (not shown).


\section{Leading-order continuum theory in arbitrary $d$}

Here we illustrate the coordinate-free derivation of the continuum theory in arbitrary space dimension $d\geq 2$ for $(d+1)$-axial strain.

\subsection{Dispersion near the nodal manifold}
\label{sec:dispnodal}

To derive the continuum theory, we require to expand the tight-binding dispersion near a point $\bs K$ on the nodal manifold. The relevant ingredient is the expansion of $f(\bs k)$ as specified in Eq.~(2) of the main text. This expansion reads
\begin{equation}
f_{\bs K} \equiv f(\bs K+ \bs q) = \bs q \cdot \partial_{\bs k} f(\bs k)|_{\bs k=\bs K} + \mathcal{O}(q^2)
\end{equation}
noting that $f(\bs K)=0$. The expansion involves a complex velocity vector $\bs c = \partial_{\bs k} f(\bs k)|_{\bs k=\bs K}$ that resides in the two-dimensional normal space of the nodal manifold at point $\bs K$. It is straightforward to show that the real and imaginary part of $\bs c$ are orthogonal to each other and hence form a basis of this normal manifold. We can thus write $\bs c = v_1 \hat{e}_1 +\im v_2 \hat{e}_2$ where $v_{1,2}$ are the (in general anisotropic) Fermi velocities (in units of $t$).

We proceed to show that $\bs c$ has constant magnitude on the entire nodal manifold. To this end, we use the explicit expression
\begin{equation}
\bs c = -\im \sum_j \exp[-\im \bs K \cdot \dhat_j] \bs \dhat_j
\end{equation}
from which we compute $|\bs c|^2$. Using the scalar-product relations for the $\dhat_j$ yields:
\begin{equation}
|\bs c|^2=(d+1) - \frac{1}{d} \sum_{l\neq j} \exp[\im \bs K \cdot\dhat_l] \exp[-\im \bs K \cdot \dhat_j]
\end{equation}
Because $\bs K$ is part of the nodal manifold, we have $\sum_j \exp[-\im \bs K \cdot \dhat_j]=0$. This enables to simplify the last term and finally gives
\begin{equation}
|\bs c| = \frac{(d+1)}{\sqrt{d}}
\end{equation}
independent of the reference point $\bs K$. We may define a global Fermi velocity as $v_F = |c|t/\sqrt{2}$ to account for the co-dimension of the manifold and make it consistent with points of isotropic dispersion. For $d=2$ this leads to the correct velocity of $\frac{3}{2}t$ for graphene \cite{Neto09}.

As we will see below, the effective pseudo-magnetic field at $\bs K$ has a strength proportional to $1/(v_1 v_2)$. Hence, we encounter a singularity if $v_1 v_2=0$, which happens if all unit vectors $\exp[-\im\bs K \cdot \dhat_j]$ are co-parallel. To identify the corresponding $\bs K$, we make the ``gauge'' choice to write $f(\bs K)$ as
\begin{equation}
f(\bs K) = 1 + \sum_{j=1}^d \exp[-\im \bs K \cdot \bs a_j] = 0
\end{equation}
where $\bs a_j$ are the Bravais lattice vectors. The condition $v_1 v_2 = 0$ requires the $\exp[-\im \bs K \cdot \bs a_j]$ each to be $\pm 1$  which is only possible in odd spatial dimensions (where there is an even number of terms summing up to zero). The corresponding momentum vector is then a linear combination of reciprocal lattice vectors, $\bs K = \sum_j d_j {\bs G}_j$ with $d_j$ either $0$ or $1/2$. Therefore the case of one velocity vanishing, $v_1 v_2=0$, can only happen at discrete time-reversal-invariant momenta $\bf K$ and only in odd spatial dimensions (e.g. at the $X$ points in $d=3$).

The Hamiltonian can locally be linearized at an arbitrary point $\bs K$ as $h_K = v_1 q_1 \sigma_x +v_2 q_2 \sigma_y$ where $q_1,q_2$ are given by projecting on the real and imaginary part of $\bs c$. This is an anisotropic Dirac theory, where -- in analogy to graphene's Dirac points -- the eigenstates pick up a Berry phase of $\pm{\rm sgn}(v_1 v_2)\pi$ when driven around the manifold in its normal space \cite{Park11}.

\subsection{Hopping amplitude and strain patterns}

As noted in the main text, we implement strain by a modification of the hopping amplitudes which are assumed to depend exponentially on the bond length, Eq.~(3) of the main text. Taking the continuum limit, the length change of a bond $(ii')$ along $\bs\delta_j$ can be expressed as $\dhat_j \cdot \underline{\underline{u}} \cdot \dhat_j$ with the strain tensor $\underline{\underline{u}}$ evaluated at the bond center. Hence
\begin{align}
t_{ii'}
&= t_0 \exp\left[ - \beta \dhat_j\cdot\underline{\underline{u}}\cdot\dhat_j \right] \\
&= t_0 (1-\beta \dhat_j\cdot\underline{\underline{u}}\cdot\dhat_j+\frac{1}{2}\beta^2(\dhat_j\cdot\underline{\underline{u}}\cdot\dhat_j)^2+\dots) \notag
\end{align}

We now specialize to the $(d+1)$-axial strain advocated in the main paper. Its form is motivated by the known displacements for triaxial and tetraxial strain in $d=2$ and $d=3$, $\bs u_{2D} = C'_2(y^2-x^2,2xy)$ and $\bs u_{3D} = C'_3(yz,xz,xy)$, respectively \cite{rachel16b,Masir13}.  The quadratic growth of $\bs u$ from the sample origin facilitates a homogeneous pseudo-magnetic field to emerge, and the fact that the corresponding strain tensor, related to the displacement field by $u_{ij}= (\partial_i u_j + \partial_j u_i)/2$, is traceless implies that there are no pseudo-electric effects in the effective Hamiltonians \cite{Neto09}. We therefore define the $(d+1)$-axial strain as
\begin{equation}
\bs u = \frac{C}{2} \sum_j (\dhat_j \cdot \bs r)^2 \dhat_j,\quad
\underline{\underline{u}} = C \sum_j (\dhat_j \cdot \bs r) (\dhat_j \circ \dhat_j)
\end{equation}
for arbitrary $d$, consistent with the above expressions for $d=2,3$ using $C'_2 = \frac{3}{4} C$ and $C'_3 = \frac{8}{3 \sqrt{3}} C$ respectively, and with a generically traceless strain tensor.
The specific form of the strain simplifies the hopping amplitudes via $\dhat_j \cdot \underline{\underline{u}} \cdot \dhat_j \rightarrow C_d \dhat_j\cdot\bs r$ where $C_d = C(1-1/d^2)$. This will be key to our analytic computations below.


\subsection{Hamiltonian and spectrum in the presence of strain}
\label{sec:strainanyd}

We now consider the hopping Hamiltonian in the presence of strain. To this end, we again expand $f(\bs k)$ about a point $\bs K$ on the nodal manifold, and take into account that the hopping matrix elements vary (slowly) in space, by also expanding in strain. For $(d+1)$-axial strain this results in
\begin{equation}
\label{fkgen}
f_{\bs K} = \sum_j e^{-\im \bs K \cdot \dhat_j} (1 - \im \bs q \cdot \dhat_j) (1 - \beta C_d \dhat_j \cdot \bs r)
\end{equation}
which depends on the two-dimensional momentum $\bs q$ relative to $\bs K$ as well as on the position $\bs r$. Noting that the zeroth-order term vanishes on the nodal manifold and keeping only linear terms, we arrive at Eq.~(12) of the main text, where the correspondence $-\im \bs q\to \partial_{\bs r}$ has been used.

This $f_{\bs K}$ renders the Schr\"odinger equation for $h_{\bs K}$ to be a two-component differential equation in real space. As usual for Landau-level problems, this is most efficiently solved algebraically by introducing ladder operators \cite{Toke}. We therefore define
\begin{equation}
b_{\bs K} = \sum_j e^{-\im \bs K \cdot \dhat_j} \dhat_j \cdot \left[\partial_{\bs r} - \beta C_d \bs r \right]\,.
\end{equation}
This operator indeed fulfills harmonic-oscillator commutation relations $[b_{\bs K}, b_{\bs K}^\dag] = {\rm const}$, as
\begin{multline}
[b_{\bs K},b_{\bs K}^\dag] = \\
\beta C_d \sum_{j,l} e^{\im \dhat_l \cdot \bs K -\im \dhat_j \cdot \bs K} \left[ (\dhat_l \cdot \partial_{\bs r}) (\dhat_j \cdot \bs r) + (\dhat_j \cdot \partial_{\bs r}) (\dhat_l \cdot \bs r) \right] \\
= 2 \beta C_d \sum_{j,l} e^{\im\dhat_l \cdot \bs K - \im \dhat_j \cdot \bs K} \dhat_l \cdot \dhat_j
\end{multline}
Using $\beta C_d = \gamma$ and the calculations for $|\bs c|, v_F$ from Sec.~\ref{sec:dispnodal} directly yields
\begin{equation}
[b_{\bs K},b_{\bs K}^\dag] = 2 \gamma \frac{(d+1)^2}{d} \equiv \omega
\end{equation}
independent of the reference point $\bs K$.

The spectrum of $h_{\bs K}^2 = t^2\left( \begin{smallmatrix} b_{\bs K}^\dag b_{\bs K}&0\\ 0&b_{\bs K}^\dag b_{\bs K} + \omega  \end{smallmatrix} \right)$ is therefore given by $n t^2 \omega$ where $n\geq0$ is a harmonic-oscillator quantum number. Taking the root yields the spectrum of $h_{\bs K}$ as quoted in Eq.~(13) in the main text,
\begin{equation}
E_n^\pm = \pm t \sqrt{n} \sqrt{2 \gamma \frac{(d+1)^2}{d}}
\label{eq:final PLL}
\end{equation}
and the corresponding eigenvectors are $(|n-1\rangle,\pm|n\rangle)$, with $|n\rangle$ the $n$-th eigenvector of the harmonic oscillator $b_{\bs K}^\dagger b_{\bs K}$. We note that this derivation does not require to explicitly introduce a pseudo-magnetic field. We can nonetheless restore the analogy to a standard relativistic Landau problem by defining an averaged dimensionless magnetic field, $\overline{B} = \gamma/2$ and using $v_F$ from Sec.~\ref{sec:dispnodal} to get
\begin{equation}
E_n^\pm = \pm \sqrt{n} \sqrt{2 \overline{B} v_F^2}\,.
\end{equation}

In summary, all points $\bs K$ on the nodal manifold generate the same relativistic PLL spectrum because their low-energy theory is effectively two-dimensional. In fact, all these low-energy theories are identical up to rotation and anisotropic rescaling. The system is therefore dimensionally reduced.

As usual for Landau-level problems, the degeneracy of the levels \eqref{eq:final PLL} scales as $L^2$ for fixed $\bs K$ where $L$ is linear system size. This degeneracy can be understood as arising from angular-momentum conservation in each two-dimensional subtheory.
This degeneracy is multiplied by the fact that the full low-energy Hamiltonian involves a summation over the nodal manifold. The number of momentum points on the nodal manifold scales as $L^{d-2}$, resulting in an overall degeneracy $\sim L^d$ of all PLLs.


\section{Continuum theory in $d=2,3$ beyond leading order}

In this section, we show details of explicit computations in $d=2,3$ where we can go beyond the leading order in the expansion in momentum and strain. We will mainly consider the 3D case, and quote the 2D result at the end of this section.


\subsection{Effective 3D Hamiltonian}

We start by writing down the explicit form of $f_{\bs K}$ \eqref{fkgen} for $\bs K = \sqrt{3}\pi/2(0,\lambda,1)$ and $\bs q = (q_x,0,q_z)$, parameterized by $\lambda$ and located on the line segment connecting the $X$ points $\sqrt{3}\pi/2(0,0,1)$ and $ \sqrt{3}\pi/2(1,0,0) \cong \sqrt{3}\pi/2(0,1,1) $ in the Brillouin zone, i.e., the blue line segment in Fig.~\ref{fig:BZ}.
\begin{multline}
f_{\bs K} =  - \frac{2}{\sqrt{3}} (\phi_\lambda (q_z-q_x) + \phi_\lambda^\ast (q_z+q_x)) \\ +\im \frac{2 \beta C_d }{\sqrt{3}} (\phi_\lambda (z-x)+ \phi_\lambda^\ast (z+x))
\end{multline}
where $\phi_\lambda = \exp[\im \frac{\pi}{2} \lambda]$ is a position-dependent complex phase factor.


\subsection{Derivation of higher orders}

For higher orders we note that by choice of $(d+1)$-axial strain we can perform expansion in orders of the combined quantity $\bs \Gamma = - \im \bs q - \beta C_d \bs r$ as the Hamiltonian can be written in a linear combination of $\exp[\bs \Gamma \cdot \dhat_j]$. Actually, we only select terms by orders of $\Gamma$, the calculations are performed for momentum and strain terms separately \cite{FN1}. Doing so keeps the semiclassical motion of wave packets bounded and enables us to perform closed algebraic calculations. The relevant terms are listed in Table~\ref{tab:s1} which are evaluated under the sum $\sum_j \exp[-\im \dhat_j \cdot \bs K] (\dots)$ so that zeroth-order terms vanish.

\renewcommand{\arraystretch}{1.5}
\begin{table}[tb]
\begin{tabular}{||c|r|p{4.5cm}||}\hline
$\mathcal{O}(\Gamma^1)$ & $-\im \dhat_j \bs q$ & first order momentum \\
& $-\gamma \dhat_j \bs r$ & first order strain \\
\hline
$\mathcal{O}(\Gamma^2)$ & $- \frac{1}{2} (\dhat_j \bs q)^2$ & second order momentum \\
& $\frac{\gamma^2}{2}(\dhat_j \bs r)^2$ & second order strain \\
& $ \im  \gamma (\dhat_j \bs q)(\dhat_j \bs r)$ & first order strain-momentum mixed term \\
\hline
$\mathcal{O}(\Gamma^3)$ & $\frac{\im}{6} (\dhat_j \bs q)^3$  & third order momentum \\
& $\frac{-\gamma^3}{6} (\dhat_j \bs r)^3$ & third order position  \\
& $\frac{-\im \gamma^2}{2}(\dhat_j \bs q)(\dhat_j \bs r)^2$ & \multirow{2}{4.5cm}{mixed first and second order terms}\\
& $\frac{\gamma}{2}(\dhat_j \bs q)^2(\dhat_j \bs r)$ &\\
\hline
\end{tabular}
\caption{Terms in the combined momentum-position expansion of $f_K$ grouped by the corresponding orders of $\Gamma$.}
\label{tab:s1}
\end{table}
\renewcommand{\arraystretch}{1}

\subsection{Second order at crossing point}

Choosing the point of reference as $K_X =  \sqrt{3} \pi/2 (0,0,1)$, i.e., the crossing point of the $x$ and $y$ nodal lines, the complex phases are
\begin{equation}
\exp[-\dhat_j \cdot \bs K_X] = \exp[\pm \pi/2] = \pm \im
\end{equation}
where the upper (lower) sign is valid for $\dhat_{3,4}$ ($\dhat_{1,2}$), respectively. From that the dispersion function $f_{\bs K}$ up to second order can be calculated straightforwardly
\begin{align}
f_X &= \underbrace{- \frac{4}{\sqrt{3}}(q_z-\im \gamma z)}_{\sim \mathcal{O}(\Gamma)} \notag \\
&+ \underbrace{\frac{4}{3}(\im q_x q_y + \gamma q_x y + \gamma q_y x - \im \gamma^2 xy )}_{\sim \mathcal{O}(\Gamma^2)}
+ \mathcal{O}(\Gamma^3)
\end{align}
Semiclassically, the first-order term describes motion along the axis orthogonal to both nodal lines (here the $z$ axis), while the second-order term moves the wave packet away from the $z$ axis. The dynamics encoded by $f_X$ can be cast into a fourth-order differential equation for the wave function that can be solved numerically.

The function $f_X$ can also be expressed using bosonic ladder operators. We define the dimensionless operators
\begin{flalign}
    & a_z = \frac{1}{\sqrt{2 \gamma}} (-q_z + \im \gamma z), \\
    & a_x = \frac{1}{\sqrt{2 \gamma}} (\im q_x +  \gamma x), \\
    & a_y = \frac{1}{\sqrt{2 \gamma}} (q_y - \im  \gamma y)
\end{flalign}
of unit measure, $[a_i,a_i^\dag]=1$, that commute with each other such that
\begin{equation}
f_X = \frac{4\sqrt{2}}{\sqrt{3}}\sqrt{\gamma} a_z + \frac{8}{3} \gamma a_x a_y.
\end{equation}
The corresponding Hamiltonian can be cast into an expansion in powers of $\sqrt\gamma$:
\begin{equation}
h_X = -t \frac{4\sqrt{2}}{\sqrt{3}} \sqrt{\gamma} \begin{pmatrix}
0 & a_z \\
a_z^\dag & 0
\end{pmatrix}
-t \frac{8}{3} \gamma \begin{pmatrix}
0 & a_x a_y \\
a_x^\dag a_y^\dag & 0
\end{pmatrix}.
\end{equation}


\subsection{Second order at any point along nodal line}

\begin{figure}[tb]
\includegraphics[scale=0.9]{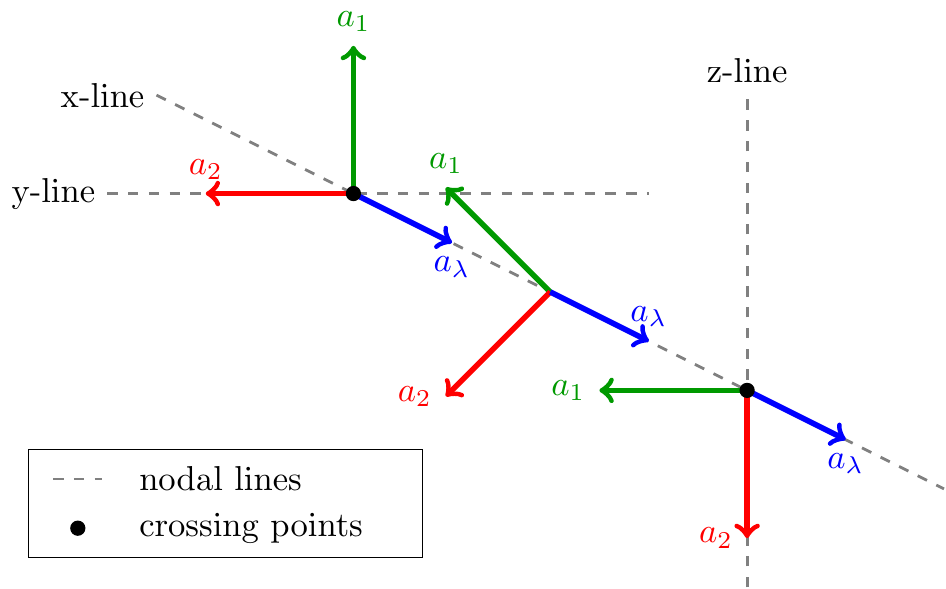}
\caption{
Rotation of the ladder operators $a_1,a_2$ along the nodal line, leaving the ladder operator $a_\lambda$ unchanged.
}
\label{fig:rotate_ladOP}
\end{figure}

Building upon the second-order expansion at the crossing point, we choose a nodal line (here along $x$) which we parameterize as $\bs K = \sqrt{3}\pi/2(\lambda,0,1)$ which gives an extra phase factor of $\varphi_\lambda = \exp[-\lambda \pi/2] $ in the expansion:
\begin{multline}
f_{\bs K} = \frac{4\sqrt{2}}{\sqrt{3}}\sqrt{\gamma} (\mathrm{Re}(\varphi_\lambda)a_z+\Im(\varphi_\lambda)a_y) + \\
+\frac{8}{3} \gamma a_x (\mathrm{Re}(\varphi_\lambda)a_y-\Im(\varphi_\lambda)a_z)\,.
\end{multline}
We can introduce rotated operators $a_1=\mathrm{Re}(\varphi_\lambda)a_z+\Im(\varphi_\lambda)a_y$ and $a_2=\mathrm{Re}(\varphi_\lambda)a_y-\Im(\varphi_\lambda)a_z$, Fig.~\ref{fig:rotate_ladOP}, then resulting in Eq.~(11) of the main text. We note that the phase factors here guarantee that at each nodal-line crossing two operators are aligned with the nodal lines. We rename $a_x$ to $a_\lambda$ to indicate that its action corresponds to a motion along the nodal line. The bosonic operators $\{a_\lambda,a_1,a_2\}$ are normalized and commute. Obviously, the operator $a_1$ is related to $b_{\bs K}$ from Sec.~\ref{sec:strainanyd} by $b_{\bs K} = \sqrt{\omega} a_1$.

The Hamiltonian at this order then reads
\begin{equation}
h_{\bs K} = -t \frac{4\sqrt{2}}{\sqrt{3}} \sqrt{\gamma} \begin{pmatrix}
0 & a_1 \\
a_1^\dag & 0
\end{pmatrix}
-t \frac{8}{3} \gamma \begin{pmatrix}
0 & a_\lambda a_2 \\
a_\lambda^\dag a_2^\dag & 0
\end{pmatrix}.
\end{equation}
Its spectrum is dictated by the matrix structure and the algebraic properties of the ladder operators; it depends only on $t\sqrt\gamma$  but not on the position on the nodal manifold.


\subsection{Perturbative analysis of second-order corrections}
\label{sec:pert2nd}
\begin{figure}[!tb]
\includegraphics[width=0.45\textwidth]{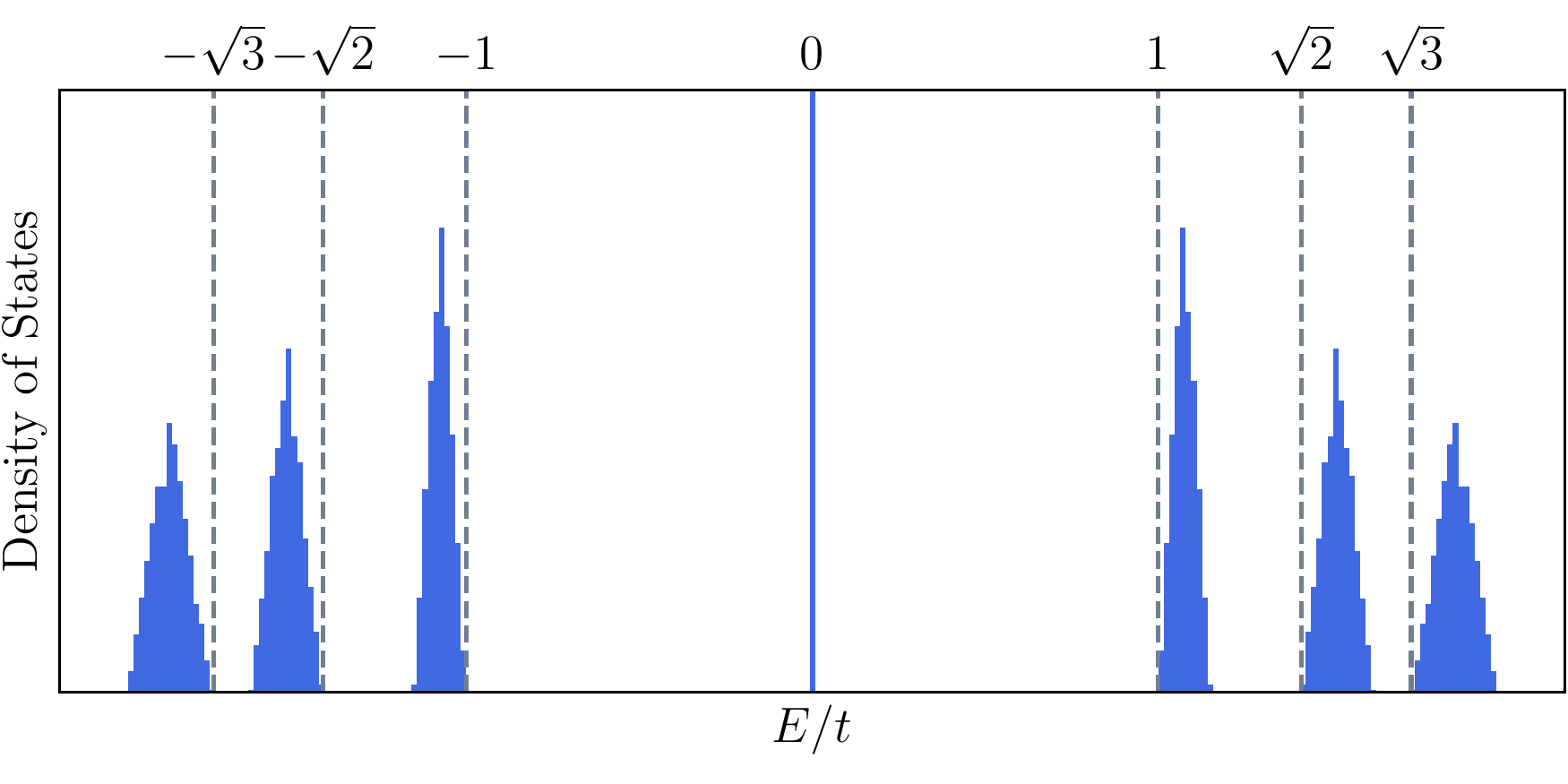}
\caption{
PLL spectrum beyond leading order, obtained from second-order perturbation theory for $t_2/t_1 = 4 \cdot 10^{-3}$ ($\gamma = 2.4 \cdot 10^{-5}$) and an excitation cutoff $N\leq 55$. The dashed lines show the unperturbed PLL energies.
}
\label{fig:2nd_order}
\end{figure}

Even tough the ladder operators $\{a_\lambda,a_1,a_2\}$ commute, the first-order and second order terms of $h_{\bs K}$ do not due to the matrix structure of the Hamiltonian. For an approximate solution, we choose an ansatz for the eigenvectors
\begin{equation}
\ket{\bs \Psi} = \begin{pmatrix}
\sum u_{nkm} \ket{n,k,m} \\ \sum d_{nkm} \ket{n,k,m}
\end{pmatrix}
\end{equation}
where $\ket{n,k,m}$ are the product eigenvectors of $N_\lambda \otimes N_1 \otimes N_2$, and $N_\lambda=a^\dagger_\lambda a_\lambda$ etc. The eigenvalue equation leads to an infinite system of linear equations for the $u_{nkm},d_{nkm}$:
\begin{flalign}
\epsilon & u_{nkm} = t_1 \sqrt{n+1} d_{n+1,km} + t_2 \sqrt{k+1}\sqrt{m+1} d_{n,k+1,m+1}, \notag \\
\epsilon & d_{nkm} = t_1 \sqrt{n} u_{n-1,km} + t_2 \sqrt{k}\sqrt{m} u_{n,k-1,m-1}
\end{flalign}
that separates into finite subsets that can be solved individually. Here, $t_1=-4\sqrt{2\gamma}/\sqrt{3}t$ and $t_2=-8\gamma/3t$ are the first-order and second-order couplings. We have performed a perturbative calculation to second order in $t_2/t_1$ on each subset of the eigenvalue equations, imposing a cutoff to the maximum number of oscillator quanta $N_i$ to enable a numerical solution. The first-order corrections in $t_2/t_1$ to the spectrum \eqref{eq:final PLL} vanish, while the second-order corrections lead to a broadening and a shift of the PLLs except for the zeroth PLL which is still perfectly degenerate, see Fig.~\ref{fig:2nd_order}.

\subsection{Second-order Hamiltonian in $d=2$}

For completeness, we specify the next-to-leading-order Hamiltonian in the case of $d=2$. We choose one of the Dirac points, $\bs K  = {4 \pi}/{3 \sqrt{3}}(0, 1)$, and like before we expand $f_{\bs K}$ in orders of $\bs \Gamma$. This yields first-order terms:
\begin{equation}
f_{\bs K} = \frac{3 \im}{2} (q_x+ \im q_y) + \frac{3 \gamma}{2} (x+\im y).
\end{equation}
As before, we define dimensionless and commuting bosonic operators:
\begin{flalign}
    & a_1 = \frac{1}{\sqrt{9 \gamma}} \left(-\frac{3 \im}{2} (q_x+ \im q_y) + \frac{3 \gamma}{2} (x+\im y) \right), \\
    & a_2 = \frac{\im}{\sqrt{4 \gamma}}\left(\phantom{+} \frac{3 \im}{2} (q_x- \im q_y) - \frac{3 \gamma}{2} (x-\im y) \right)
\end{flalign}
to express the Hamiltonian as
\begin{equation}
h_{\bs K} = t  \sqrt{9\gamma} \begin{pmatrix}
0 & a_1 \\
a_1^\dag & 0
\end{pmatrix}
+ t \frac{3}{2} \gamma \begin{pmatrix}
0 & (a_2)^2 \\
(a_2^\dag)^2 & 0
\end{pmatrix}
\end{equation}
which has the same matrix structure as the effective Hamiltonian in $d=3$, but features only two (instead of three) bosonic operator. Therefore one can apply the same perturbative method presented in Sec.~\ref{sec:pert2nd} to derive a similar broadening for the pseudo-Landau levels as illustrated in Fig.~\ref{fig:2nd_order}. Furthermore we can immediately infer that the leading-order PLL spectrum is given by $E_n = \pm t \sqrt{n} \sqrt{9 \gamma}$, consistent with Eq.~\eqref{eq:final PLL}.


\section{Semiclassical motion in $d=3$}

One key feature near the Fermi surface is the description of the problem as an effective two-dimensional subtheory. This can only be valid if the subtheories for different points on the Fermi surface are not interacting with each other. We validate this by showing that the semiclassical motion in leading order does not leave its subtheory slice.


\subsection{Motion near the nodal line}

The unstrained Hamiltonian is solved by Bloch states $\Psi_{A,\bs k}, \Psi_{B,\bs k}$ where $A,B$ denote the sublattice, which combine to the eigenstates
\begin{equation}
\Psi_{\bs k}^{(\pm)} = \Psi_{A,\bs k} \mp \frac{f(\bs k)^\ast}{|f(\bs k)|}\Psi_{B,\bs k}
\end{equation}
with energy $E^\pm(\bs k) = |f(\bs k)|$. A wave packet in the upper band can be constructed
\begin{equation}
\Psi_{\rm WP} = \int d^3k u(\bs k) \Psi_{\bs k}^+
\end{equation}
with a modulation $u(k)$ that results in a sharp wave packet relative to the change in strain. Then the motion of the wave packet is approximated by the motion of the real and momentum space means $\bar{\bs r} = \langle \bs r \rangle , \bar{\bs k} = \langle \bs k \rangle  $ which obey (to leading order) Hamilton's equations of motion.
\begin{align}
\frac{d}{dt} \bar{\bs r}  &= \phantom{-}\frac{\partial E}{\partial \bs k}, \\
\frac{d}{dt} \bar{\bs k}  &= -\frac{\partial E}{\partial \bs r}.
\end{align}
We chose an arbitrary point $\bs K_\lambda$ on the nodal line in $z$ direction, parameterized by $\lambda$, and express the energy in proximity to the nodal line
\begin{multline}
(E/t)^2 = \left(\frac{4}{\sqrt{3}}v_x (\lambda) q_x+\frac{4 \gamma}{\sqrt{3}} v_y(\lambda) y \right)^2 \\+\left(\frac{4}{\sqrt{3}}v_y (\lambda) q_y-\frac{4 \gamma}{\sqrt{3}} v_x (\lambda) x \right)^2
\end{multline}
where $\bs q$ is the momentum relative to $\bs K_\lambda$. The dimensionless Fermi velocities fulfill $v_x(\lambda)^2+v_y(\lambda)^2 = 1$, for the remainder we omit the explicit dependence on the line position $\lambda$. The coordinates and time $\tau$ can be rescaled to absorb all prefactors $\bs r \rightarrow \bs r \sqrt{3}/4 \gamma, \bs q \rightarrow \bs q \sqrt{3}/4, \tau \rightarrow \tau  (3 E)/(t^2 16 \gamma)$ where $E$ is the constant energy of the classical orbit. This finally yields the dimensionless equations of motion
\begin{IEEEeqnarray}{lcl}
\begin{pmatrix}
\dot{x} \\ \dot{y} \\ \dot{z}
\end{pmatrix} &=& \begin{pmatrix}
(v_x q_x+ v_y y)v_x \\
(v_y q_y-v_x x ) v_y \\
0
\end{pmatrix} \\
\begin{pmatrix}
\dot{q}_x \\ \dot{q}_y \\ \dot{q}_z
\end{pmatrix} &=& \begin{pmatrix}
(v_y q_y-v_x x )v_x \\
 -(v_x q_x+ v_y y)v_y \\
 0
\end{pmatrix}
\end{IEEEeqnarray}
specific to $\bs K_\lambda$. The in-plane orbits are
\begin{equation}
\bs r = \phantom{-} A_1 (v_x \cos \tau, - v_y \sin \tau) + A_2 (v_x \sin \tau, v_y \cos \tau)
\end{equation}
and
\begin{equation}
\bs q = -A_1 (v_x \sin \tau, v_y \cos \tau) + A_2 (v_x \cos \tau, -v_y \sin \tau)
\end{equation}
that obey the relation $E=t\sqrt{A_1^2+A_2^2}$ and whose centers can be mutually shifted by two linearly independent vectors in phase space. This adds a degeneracy per subtheory of $L^2$ as can be seen from the standard 2D Landau problem with finite linear system size.

\subsection{Sommerfeld requantization}

As the semiclassical orbits are closed in phase space it is possible to impose quantization conditions on the enclosed phase space volume \cite{Xiao10}. The current coordinates $(x,y,q_x,q_y)$ can be transformed into two independent pairs with the canonical transformation
\begin{equation}
\begin{pmatrix}
\omega \\ p_\omega \\ \gamma   \\ p_\gamma
\end{pmatrix}
=\begin{pmatrix}
0 & v_x & v_y & 0 \\
-v_x & 0 & 0 & v_y \\
-v_y & 0 & 0 & -v_x \\
0 & -v_y & v_x& 0
\end{pmatrix}
\begin{pmatrix}
x \\ q_x \\ y  \\ q_y
\end{pmatrix}
\end{equation}
which leaves the Hamiltonian as
\begin{equation}
H(\omega,\gamma, p_\omega, p_\gamma) = t \sqrt{\omega^2+p_\omega^2}
\end{equation}
and only dependent on $(\omega,p_\omega)$. We apply a simple Sommerfeld's quantization rule, $\int p_\omega d\omega = (n+\frac{1}{2} - \frac{\Gamma}{2\pi}) h$, where $\Gamma$ is the Berry phase of the closed semiclassical orbit, also we set the Maslov index to $\frac{1}{2}$ as $H^2$ is an harmonic oscillator. We previously argued that the Berry phase around the nodal line is either $\pm \pi$, but the semi-classical orbit selects the positive Berry phase and cancels the constant terms \cite{Xiao10, Fuchs10}.
Doing so leads (after rescaling the coordinates) to the positive branch of Eq.~\eqref{eq:final PLL}
\begin{equation}
E = t \sqrt{\frac{32 \gamma}{3}} \sqrt{n}
\end{equation}
as a wave packet of negative energy would lead to the other branch.


\section{Aspects of gauge invariance}

This section is to summarize the conceptual difference between magnetic-field induced Landau levels and strain-induced PLLs concerning gauge invariance.

In the context of physical electromagnetic fields, gauge invariance refers to different mathematical descriptions of the same physics, i.e., a given form of the vector potential can be transformed into a different one by a gauge transformation yielding identical observables. This is very different when discussing effects of strain on the single-particle dynamics. Here, strain enters as an effective vector potential, dubbed pseudo-vector potential, from which a pseudo-magnetic field can be computed. Now, going from one pseudo-vector potential to another by a pseudo-gauge transformation corresponds to changing the strain pattern. To leading order in the continuum theory, this may still yield the same pseudo-magnetic field as before, but beyond this leading order this does not hold. In addition, strain can also induce pseudo-electric fields (and more complicated higher-order terms) which are not invariant under pseudo-gauge transformations. Both aspects can be seen in numerical lattice calculations where different choices strain, formally connected by a pseudo-gauge transformation, lead to different single-particle spectra \cite{vozmediano10,Masir13,sturla13,castro17,vojta23}.

The triaxial strain often used for graphene corresponds to symmetric gauge and have the property that pseudo-electric fields vanish identically. The same applies to the $(d+1)$-axial strain proposed here, which yields a homogeneous pseudo-magnetic field in symmetric gauge in any of the two-dimensional subtheories.


\section{Limits to the strain amplitude}

The inhomogeneous $(d+1)$-axial displacement field as specified in Eq.~(4) of the main paper has the property that it scales as $r^2$ where $r$ is the distance of a sample point from its center; therefore the local change of bond lengths scales as $r$. Mechanical stability requires that this change is small compared to the original lattice spacing. For a sample of linear size $L$, the maximum allowed $C$ therefore scales as $1/L$. This also implies that the thermodynamic limit, $L\to\infty$, cannot be taken at fixed $C$, i.e., fixed pseudomagnetic field. It is, however, possible to take a combined limit $L\to\infty$, $C\to 0$ with $CL$ kept fixed \cite{rachel16b}.

In practical realizations, the displacement field will scale as $r^2$ only in the interior of the sample, such that the pseudomagnetic field display inhomogeneities near the sample edges, similar to what has been discussed for graphene \cite{peeters13}. Alternatively, periodic schemes of strain have been proposed.
